\documentclass[a4paper,11pt]{article}
\usepackage[T1]{fontenc}
\usepackage[utf8]{inputenc}
\usepackage{lmodern}
\usepackage{microtype}
\usepackage[top=23mm,bottom=24mm,left=24mm,right=24mm]{geometry}
\usepackage{graphicx}
\usepackage{caption}
\usepackage{pdflscape}
\usepackage{placeins}
\usepackage{enumitem}
\usepackage{titlesec}
\usepackage{xurl}
\usepackage[numbers,sort&compress]{natbib}

\usepackage[hidelinks,unicode]{hyperref}
\graphicspath{{figures/}}
\titleformat{\section}{\large\bfseries}{\thesection}{0.65em}{}
\titlespacing*{\section}{0pt}{2.1ex plus 0.6ex minus 0.2ex}{0.8ex plus 0.2ex}
\date{}
\hypersetup{pdftitle={Predator-associated cues promote host riding, and coupling to mobile hosts improves survival in an epizoic limpet},pdfauthor={Ryo Nakayama; Katsushi Kagaya; Keita Harada; Tomoyuki Nakano},pdfsubject={Preprint manuscript; 16 November 2026; proofreading revision},pdfkeywords={}}
\begin{document}
\begin{center}
{\fontsize{16}{20}\selectfont\bfseries Predator-associated cues promote host riding, and coupling to mobile hosts improves survival in an epizoic limpet\par}
\vspace{1em}
{\normalsize Ryo Nakayama\textsuperscript{1,3,*}, Katsushi Kagaya\textsuperscript{1,5,*}, Keita Harada\textsuperscript{2,4}, Tomoyuki Nakano\textsuperscript{1}\par}\vspace{0.65em}
{\small
\textsuperscript{1}Seto Marine Biological Laboratory, Field Science Education and Research Center, Kyoto University, Shirahama, Wakayama, Japan\par
\textsuperscript{2}Shirahama Aquarium, Kyoto University, Shirahama, Wakayama, Japan\par
\textsuperscript{3}Current address: Fisheries Research Institute, Aomori Prefectural Industrial Technology Research Center, Hiranai, Aomori, Japan\par
\textsuperscript{4}Current address: Green Thermal Wakayama Company, Limited, Kamitonda, Wakayama, Japan\par
\textsuperscript{5}Current address: Faculty of Engineering, Kitami Institute of Technology, Kitami, Hokkaido, Japan\par
}\vspace{0.55em}
{\small \textsuperscript{*}Ryo Nakayama and Katsushi Kagaya contributed equally to this work.\par}
{\small Corresponding author: kkagaya@mail.kitami-it.ac.jp\par}
\end{center}

\vspace{0.3em}
\begin{abstract}
\noindent
For small animals, a host with multiscale movements may be difficult to intercept or predict. Boarding turns that moving target into a source of locomotion. The tegulid snail \textit{Tegula nigerrima} was previously shown to exhibit statistical signatures consistent with near-critical locomotor dynamics over limited ranges. Taking this prior characterisation as our starting point, we examined how the facultative epizoic limpet \textit{Lottia tenuisculpta} approaches and boards this host, and whether host mobility improves rider survival. Crab-associated waterborne cues increased the proportion of riders (19 of 80 without cues versus 42 of 76 with cues) and increased visible host--limpet distance closure. Paired trajectories of eventual riders captured projected movement by both animals toward one another during a nominal ten-minute window ending at the reference frame for movement analysis. After normalisation by each animal's shell size, the limpet's projected contribution exceeded the host's in 22 of 58 pairs. An exploratory analysis of a complementary assay compared each limpet's cumulative path length on an assigned surface with its no-mucus baseline at a common observation time. The log-transformed ratios varied less among individuals in the gastropod-mucus groups than in the no-mucus group. Across the observed frequency range, a matched-trial comparison estimated a 1.21-fold cue-present to cue-absent amplitude ratio (95\% recording-date bootstrap interval: 0.97--1.59), while the between-treatment difference in high-frequency spectral slope remained uncertain. Limpet survival was higher on mobile than on fixed hosts, with host shells occupied in both treatments (posterior median hazard ratio, fixed versus mobile: 2.1). These findings connect the behavioural formation of host--rider coupling to a survival benefit from host mobility, providing a case study of how one animal gains access to the multiscale locomotion of another.

\end{abstract}
\section{Introduction}
Small animals often use larger organisms to overcome limitations on movement and habitat access, or to reduce exposure to risk. In marine systems, epizoic animals live on the surfaces of larger hosts \cite{wahl1989,wahl2008}. Host riding can turn these hosts into temporary or persistent mobile platforms, as illustrated by phoresy and hitchhiking \cite{houck1991,kano2009}. Host use can reflect local adaptation, including differences in host preference and performance \cite{sotka2005}. The behaviour of hosts and epibionts can also influence habitat choice and the evolution of specialisation among shell-dwelling organisms \cite{vermeij2020}. Using a mobile host, however, poses a dynamical problem: a prospective rider must locate, approach, and board a host that may move faster than itself, then remain coupled as the host continues to move. Under predation risk, this problem becomes urgent: failure to board leaves the smaller animal exposed, whereas successful coupling may convert the same moving host into a mobile refuge.

The intertidal limpet \textit{Lottia tenuisculpta} provides a system in which to examine this problem. It uses several gastropod hosts, including the tegulid snail \textit{Tegula nigerrima} \cite{nakayama2020,nakayama2022}, and has been identified as a facultative epizoite with broad host use \cite{nakayama2022}. Host use is concentrated among small limpets; individuals larger than about 7 mm in shell length are rarely found on tegulid hosts in the field \cite{nakayama2020}. These studies established the limpet's facultative host use. Two questions remain: how does predation risk influence approach and boarding, and does host mobility itself improve rider survival? These questions concern both the behavioural formation of host--rider coupling and its consequences for survival.

In this system, the prospective rider encounters a host whose locomotion varies across timescales. Exploratory locomotion of \textit{T. nigerrima} was previously shown to exhibit 1/f-like speed fluctuations and, over limited ranges, multiple power laws and a scaling relation used to assess criticality \cite{kagaya2025}. These findings are consistent with near-critical locomotor dynamics, here described as critical-like. This previous characterisation provides the dynamical starting point for the present study, which focuses on how a rider forms an association with this mobile host and what it gains from doing so. Such multiscale motion may make the host difficult for a smaller animal to intercept or predict. Once aboard, however, the rider is carried by the host, and the same motion may make the coupled pair more difficult for a predator to track or capture. Host riding thus creates a route from approaching a complex moving target to participating in its dynamics.

The transition to host riding connects this system to a broader question: how do animals interact adaptively with environments that vary across scales? Ecologically relevant patterns and disturbances can span multiple spatial and temporal scales, requiring nervous and behavioural systems that remain sensitive while allowing flexible responses. One influential account proposes that neural systems gain such flexibility near criticality, where correlations across scales can support a large dynamical repertoire, sensitivity, information transmission, and transitions between behavioural states \cite{chialvo2010,munoz2018,shew2013}. Broader syntheses have proposed criticality as a balance between robustness and flexibility in living systems \cite{munoz2018}. For a rider, however, movement is generated partly by the host, creating a potential route to useful dynamics outside the rider's own body. This raises a complementary question: can an animal gain functional benefits by coupling to movement generated by another organism? Host riding provides a concrete ecological setting in which to examine this possibility for a host with previously characterised critical-like locomotion.

How might a limpet establish this coupling with a moving host? Gastropod pedal mucus offers a plausible route. Beyond their roles in lubrication and adhesion, mucus trails can serve as persistent external sources of chemical and contact information, guide movement, and mediate homing, aggregation, mate searching, and predator--prey interactions \cite{wada2026,ng2013}. A moving snail's trail provides a local record of its recent movement. A limpet could use this record to guide its search through local responses, potentially reducing the need for long-range visual tracking or prediction of the host's full trajectory. Predation risk can alter prey behavioural decisions and generate nonlethal effects \cite{lima1990,lima1998,werner2003,preisser2005}, while waterborne chemical cues can elicit antipredator responses in marine gastropods \cite{kats1998,jacobsen2004,ojima2013}. In this system, such cues may increase the urgency or probability of approaching and boarding a host. Together, these ideas suggest a possible mucus-linked route into host--rider coupling.

With the host's locomotion already characterised in previous work, we focus here on the rider's approach, the formation of the host--rider association, and the survival value of host mobility. We show that predator cues promote host riding and closer approach in \textit{L. tenuisculpta}, and that host mobility increases rider survival. We combined host-riding assays with paired-trajectory decomposition to quantify visible host--limpet distance closure and compare the two animals' projected displacements toward one another after normalising each component by the corresponding animal's shell size. A mobile-versus-fixed host experiment tested the survival value of host mobility while retaining shell occupancy in both treatments. A complementary mucus-exposure assay examined changes in limpet movement to investigate mucus as a candidate sensory route into coupling. Together with the previously characterised critical-like locomotion of \textit{T. nigerrima} \cite{kagaya2025}, these results link the formation of an interspecific coupling to a survival benefit from host-generated movement.

\section{Materials and methods}
\begin{figure}[!t]
\centering
\includegraphics[width=112mm]{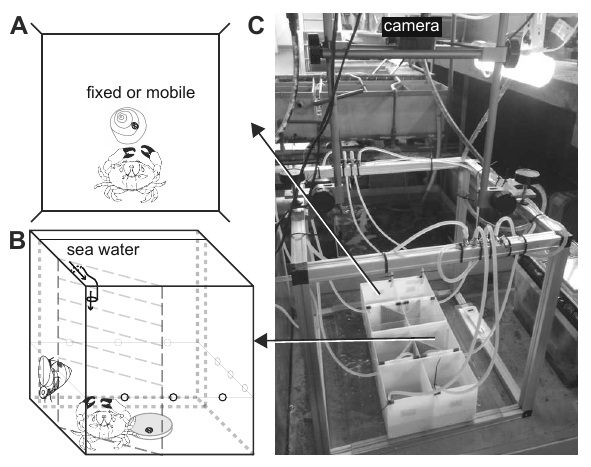}
\caption{Experimental design and tracking setup. (A) Survival assay for \textit{Lottia tenuisculpta} riding on the shell of \textit{Tegula nigerrima}. The host snail was either fixed to the chamber floor or allowed to remain mobile, whereas the limpet was never restrained. (B) Riding assay for the response of \textit{L. tenuisculpta} to predator-associated waterborne cues. In the predator-cue-present treatment, a crab was confined in a separate compartment behind a perforated partition, allowing waterborne cues to enter the test compartment while preventing contact; in the predator-cue-absent treatment, the corresponding compartment was empty. (C) Overhead tracking setup used to record locomotion and interaction trajectories. The crab and \textit{T. nigerrima} host-snail illustrations in panels A and B were created by Mari Nakano and are licensed under CC BY 4.0.}\label{fig:1}
\end{figure}

We used the limpet \textit{Lottia tenuisculpta} as the rider species, the snail \textit{Tegula nigerrima} as the host species, and the carnivorous crab \textit{Leptodius affinis} as the predator species and the source of predator cues. Previous work reported no apparent effect of crab presence on host-snail locomotion~\cite{kagaya2025}. Animals for the riding and survival assays were collected near the Seto Marine Biological Laboratory, Shirahama, Japan, between August and September 2019. Limpets used in these assays were smaller than 7 mm in shell length (2.35--5.29 mm). Trials were recorded overnight with time-lapse imaging at 20-s intervals. Coordinates used for locomotion and paired-trajectory analyses were extracted from the time-lapse images using markerless tracking with DeepLabCut version 3.0.0 rc10 \cite{mathis2018}. The survival and riding layouts and the overhead tracking setup are shown in Fig.~\ref{fig:1}.

In the host-riding assay (Fig.~\ref{fig:1}B), each chamber contained one host snail and four marked limpets. In the predator-cue-present treatment, a crab was confined in a separate compartment behind a perforated barrier, allowing waterborne cues to enter the test compartment while preventing contact. In the predator-cue-absent treatment, the corresponding compartment was empty. Visual contact was restricted by the perforated barrier, but chemical and vibrational stimuli were not experimentally separated. Twenty overnight trial sets (40 chambers) were initiated; one chamber was excluded because complete riding annotation was unavailable. The final riding dataset therefore comprised 156 limpets from 39 chamber-level observations: 80 predator-cue-absent and 76 predator-cue-present individuals. A limpet was scored as a rider if it rode on the host snail at least once during the overnight observation period. We analysed riding with a Bernoulli-logit model including predator cue presence as a fixed effect and a normally distributed random intercept for chamber. Reported riding probabilities were calculated with the chamber effect set to zero, and the predator cue odds ratio was the exponentiated fixed-effect coefficient. Figure~\ref{fig:riding}A provides a descriptive overview of recorded track-endpoint times for eventual riders. The model instead addresses whether riding occurred at least once during the overnight observation period.

The tracking input comprised 195 individual trajectories (156 limpets and 39 host snails) from 39 chambers. Chamber recordings spanned approximately 16 h at 20-s intervals; individual trajectory lengths and coordinate availability varied. Across all trajectories, the input contained 557,591 animal--frame records, of which 430,826 had non-missing $x$--$y$ coordinates. Tracking outputs were manually corrected before analysis, as described in the electronic supplementary material. Sample sizes and observation windows for individual analyses are reported separately.

For visible paired trajectories, host and limpet coordinates were paired only when both animals were visible in the same chamber and image frame. Host--limpet distance was converted to centimetres using the calibration 10 cm / 500 px = 0.02 cm per pixel. For each limpet, time 0 was the final frame in which both animals were visible. This alignment defines the end of the visible paired trajectory, which may differ from the moment of boarding. Distance closure over the final visible five minutes was calculated as the median distance in the first three samples of the window minus the median distance in the final three samples; positive values indicate decreasing host--limpet distance. This analysis included 148 limpets with at least six visible paired frames in the window (74 per predator cue treatment across 39 chambers). We denote the difference in mean distance closure between treatments by $\Delta D$ (cue present minus cue absent). To estimate its uncertainty, we resampled chambers with replacement within each treatment, retained all eligible limpets in each selected chamber, and used the 10th and 90th percentiles from 5,000 bootstrap replicates. 

For eventual riders, we decomposed each animal's movement toward the other within a nominal ten-minute window ending at a reference frame for movement analysis. The reference was the final complete limpet-track record in the original analysis input, with both coordinates and speed available. We retained paired coordinates up to and including that frame. This reference can precede the last visible paired frame used for distance closure; it was defined from track completeness rather than an independent annotation of boarding. Segments were selected by their later endpoint, including endpoints from $-10$ to 0 min. A segment ending exactly at $-10$ min can therefore begin 20 s earlier, giving at most 10 min 20 s of movement in the nominal ten-minute window. We required at least 15 valid 20-s movement segments formed from consecutive paired frames. Three of the 61 riders did not meet this criterion: 21-2 and 35-3 in the predator-cue-present treatment had 3 and 4 valid segments, respectively, and 40-2 in the predator-cue-absent treatment had 4. This left 58 riders in the ten-minute decomposition. The ten-minute component trajectories and observed values are shown in Fig. S3; the corresponding five-minute decomposition was retained as a shorter-window check in the electronic supplementary text.

For each 20-s interval, we projected each animal's movement onto the host--limpet axis defined at the start of the interval. Positive values indicate displacement toward the other animal along that axis. These projections describe each animal's contribution to movement along the axis; their sum is not, in general, the exact finite-step change in Euclidean separation. To compare animals of different sizes, we divided limpet movement toward the host by limpet shell length, and host-snail movement toward the limpet by host shell width. We refer to these as the size-normalised projected contributions. Because the normalised values were asymmetric and right-skewed, Fig.~\ref{fig:riding}C overlays the observed values with posterior predictive densities from a descriptive rider-level skew-$t$ model (Fig. S4). Figure S1 provides a descriptive overview of all visible host-snail and limpet tracks by chamber and predator cue treatment, with lines broken across missing frames or implausible jumps. It serves to show the individual paths rather than provide an additional inferential test.

The survival assay (Fig.~\ref{fig:1}A) compared limpets riding on mobile hosts with those riding on fixed hosts. Each observation involved one limpet on a living host snail and one crab in a test chamber. Each numbered trial paired one observation from each host treatment in separate chambers, with a different crab in each. A small amount of instant adhesive was applied to the underside of the host in both treatments. Fixed hosts were secured to the chamber floor; for mobile hosts, the adhesive dried without securing the host to any surface. The limpet itself was not restrained. The endpoint was predation on the limpet. Forty observations (20 treatment pairs) were initiated. Nine observations had no valid endpoint: the host lost fixation in four fixed-host trials, and the limpet detached from its host in five trials (two fixed and three mobile). All nine were excluded rather than right-censored, leaving 31 valid observations (17 mobile and 14 fixed). Limpets that remained on their hosts and were not eaten by the end of observation were right-censored. We analysed time-to-predation data in a discrete-time survival framework \cite{singer1993}, implementing a log-link hazard model with host treatment as the predictor and retaining right-censored observations. The reported hazard ratio is the fixed-to-mobile ratio of conditional one-minute predation probabilities; these hazards were constant within intervals between distinct observation times (electronic supplementary material).

To examine a candidate mucus-mediated route into coupling, we conducted a complementary mucus-exposure assay with individually identified \textit{L. tenuisculpta} in Hiranai, Aomori, Japan, in 2025. Each limpet was first observed on a no-mucus surface and later assigned to a no-mucus, \textit{Tegula} mucus, or \textit{Monodonta} mucus condition. \textit{Tegula rustica}, another species in the same genus as \textit{T. nigerrima}, served as a mucus-donor proxy, whereas \textit{Monodonta confusa} provided mucus from a sympatric gastropod that is rarely used as a host. Baseline and assigned runs were matched by limpet identifier. For 22 individuals, both runs included an observation at 200 min. At this common time point, we calculated $R_i$, the ratio of cumulative path length in the assigned run to that in the baseline run for limpet $i$. We modelled $\log R_i$ using a normal distribution with a separate mean $\mu_g$ and standard deviation for each treatment $g$. Thus $\exp(\mu_g)$ is the population median of $R_i$ in treatment $g$. The no-mucus reassignment group provided a repeated-measurement comparison. Husbandry details, exclusions, model priors, and sampling diagnostics are given in the electronic supplementary material.

For a complementary description of predator-cue-associated limpet locomotion, we calculated frame-to-frame movement within each individual track, treating the first value as missing because it has no preceding position from that individual. We calculated amplitude spectra from mean-centred series prepared from each limpet's longest available runs without missing movement values. We defined amplitude as the modulus of the Fourier transform divided by the number of samples in the retained series. This yields an amplitude spectrum rather than a power spectrum. To summarise amplitude across the observed positive-frequency range, we divided log-frequency into 16 equal-width bins. Within each bin, we calculated a median log-amplitude for each limpet, averaged individuals within chamber, and compared cue-present and cue-absent chambers within each complete trial set. We averaged available paired contrasts within each bin and then averaged the 16 bin contrasts equally. Recording dates, which each contained two trial sets, were resampled with all associated trials, chambers, animals and frequencies together. This analysis estimates an average contrast across the observed frequency range, not separate effects at every frequency. The pooled points in Fig.~\ref{fig:spectrum} are descriptive. We estimated an individual log--log slope over the high-frequency range ($\log_{10}$ frequency in Hz $\geq-3.1$), then analysed those estimates with a chamber-level hierarchical normal model that incorporated their standard errors (electronic supplementary material, Section 6). Neither inferential analysis treats Fourier ordinates as independent animals. The slope model accounts for chamber grouping but not dependence within trial sets or recording dates.

Probabilistic models were implemented in the probabilistic programming language Stan \cite{carpenter2017,stan238}, and posterior samples were drawn using Hamiltonian Monte Carlo. The corrected locomotor-spectrum inference was independently checked by analytic marginalisation and numerical integration of the same model. Additional model details, posterior checks, the full mucus-exposure analysis, and details of the locomotor-spectrum calculation are provided in the electronic supplementary material.

\section{Results}
We first examined how predator cues affected limpet behaviour and the formation of host--rider coupling, then quantified the animals' respective contributions to approach. A complementary mucus assay examined a candidate sensory route into this association. Finally, a separate mobile--fixed host comparison tested whether host mobility benefited limpets once they occupied a host shell.

\begin{figure}[htbp]
\centering
\includegraphics[width=\linewidth]{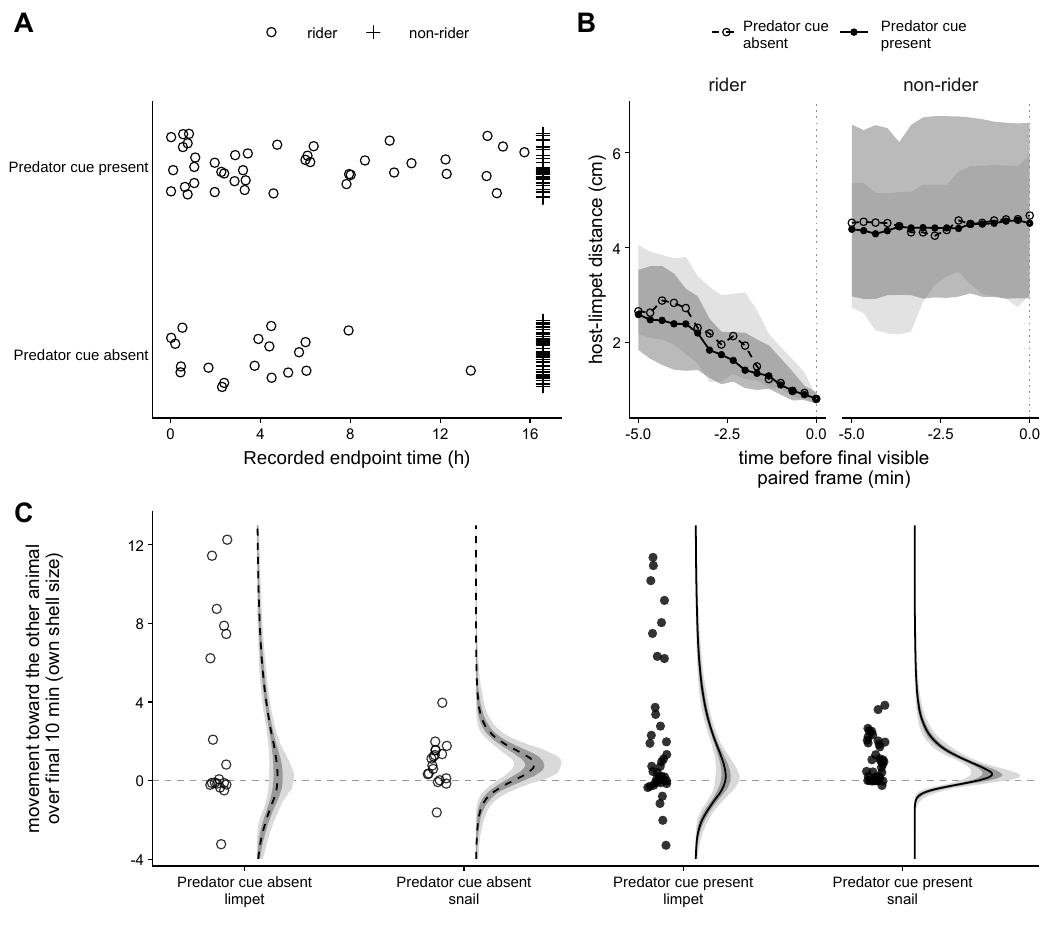}
\caption{Host riding and visible approach in the predator-cue-absent and predator-cue-present treatments. Predator cue treatments are distinguished by row labels in (A), and by open circles and dashed lines (predator cue absent) versus filled circles and solid lines (predator cue present) in (B,C). (A) Recorded track-endpoint times for eventual riders in predator-cue-absent and predator-cue-present treatments. Open circles denote individuals that rode at least once and are positioned by their final complete limpet-track record; plus signs beyond the observation endpoint indicate non-riders. (B) Host--limpet distance during the final visible five minutes, aligned to each limpet's final visible paired frame at time 0 and shown separately for eventual riders and non-riders. Lines and symbols show medians; light and dark grey ribbons show the 25--75\% ranges for predator-cue-absent and predator-cue-present animals, respectively. (C) Body-size-normalised movement toward the other animal in the nominal ten-minute window ending at each rider's movement-analysis reference frame, defined in the Methods. Points show observed values. Dashed and solid curves show posterior median predictive densities for predator-cue-absent and predator-cue-present animals, respectively; dark and light grey bands show pointwise 50\% and 90\% posterior intervals for those densities from a rider-level skew-$t$ model. Limpet movement is scaled by limpet shell length and host-snail movement by host shell width; positive values indicate cumulative projected displacement toward the other animal.}\label{fig:riding}
\end{figure}

Predator cues increased host riding, indicating a change in limpet behaviour that promoted host--rider coupling. In raw counts, 19 of 80 predator-cue-absent limpets rode on the host snail during the observation window, whereas 42 of 76 predator-cue-present limpets did so (Fig.~\ref{fig:riding}A). The chamber-level hierarchical model estimated posterior median riding probabilities of 0.215 in the predator-cue-absent treatment and 0.552 in the predator-cue-present treatment. The posterior median odds ratio for the predator cue effect was 4.534, and its 10--90\% posterior interval lay entirely above 1.

\begin{figure}[htbp]
\centering
\includegraphics[width=\linewidth]{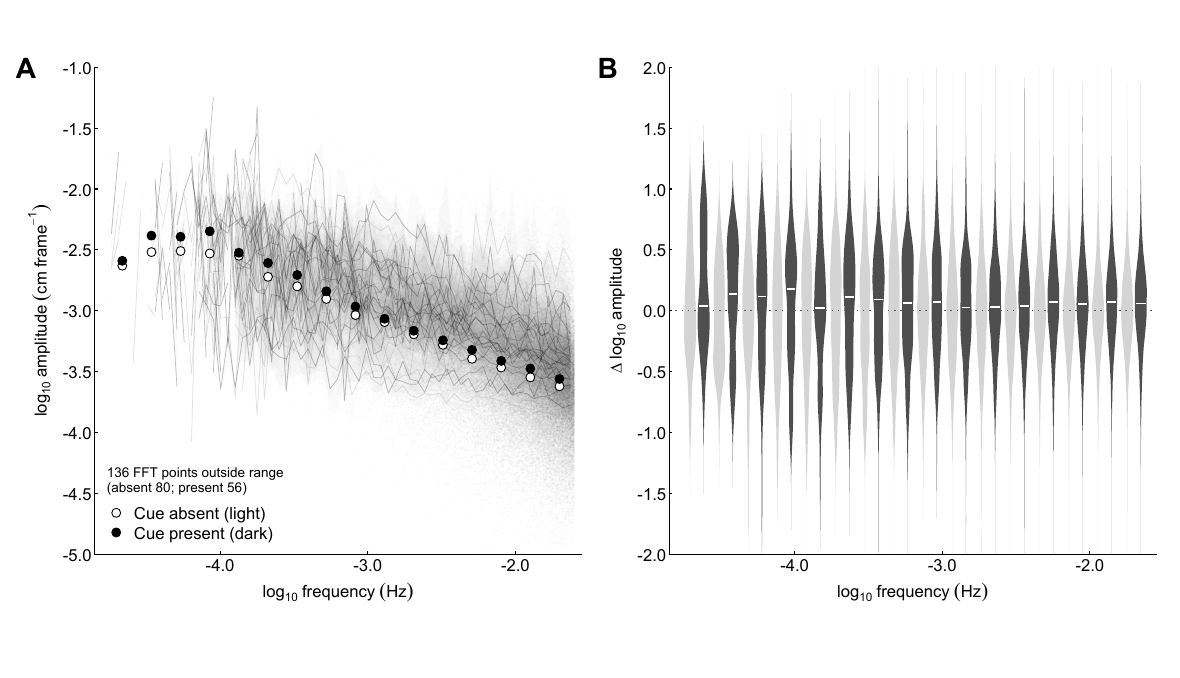}
\caption{Limpet locomotor amplitude spectra and within-frequency distributions in predator-cue-absent and predator-cue-present treatments. Spectra use each limpet's longest valid, mean-centred movement series; amplitudes are Fourier moduli divided by sample count (cm frame$^{-1}$), at 0.05 Hz sampling. The 108,708 positive-frequency, positive-amplitude ordinates came from 155 limpets in 39 chambers. (A) Open and filled circles show pooled log-amplitude medians in 16 equally spaced log-frequency bins. Faint points show Fourier ordinates; thin curves and bands show chamber medians and interquartile ranges of individual medians in 64 finer bins. The plotted range omits 136 ordinates below $-5$ (80 cue absent; 56 cue present) without excluding them from calculations. (B) Within each 16-bin interval, the pooled cue-absent median was subtracted from both groups. Light and dark shapes show kernel densities, with a common bandwidth and density-to-width scale within each bin but separate scales across bins. The dotted line is zero; white segments mark pooled cue-present medians. The displayed vertical range is $-2$ to $+2$; 17 underlying ordinates lie below and one above it, but all contribute to the densities and medians. Densities are descriptive, not confidence intervals. Inferential analyses are described in the Methods.}\label{fig:spectrum}
\end{figure}

The pooled cue-present median amplitude was higher in each of the 16 observed frequency bins (Fig.~\ref{fig:spectrum}A,B), a descriptive comparison that weights animals by their numbers of Fourier ordinates. For 19 complete trial sets, the paired difference averaged over the 16 bins was 0.082 log$_{10}$ amplitude units, corresponding to a ratio of 1.21 (80\% recording-date bootstrap interval: 1.03--1.43; 95\% interval: 0.97--1.59). The 95\% interval includes no difference. The treatment difference in the median of individual bin medians was positive in 14 of 16 bins. A fixed-512-frame sensitivity analysis covered only 12 bins and gave a ratio of 1.04 (80\% interval 0.79--1.39), with fewer animals; this check changes both sampling and frequency coverage (electronic supplementary material). The estimated treatment contrast in high-frequency amplitude-spectrum slope was small and uncertain (electronic supplementary material, Section 6). The hierarchical analysis included 79 predator-cue-absent and 74 predator-cue-present limpets. Posterior median slopes were -0.445 (10--90\% posterior interval: -0.475 to -0.414) and -0.428 (-0.459 to -0.397), respectively. The posterior median predator-cue-associated difference was 0.017, with a 10--90\% interval of -0.027 to 0.060. This comparison provides limited evidence for a treatment-level shift in the high-frequency slope.

Visible paired trajectories also showed greater host--limpet distance closure in the presence of predator cues. This analysis addressed how encounters formed by measuring the animals' approach during the final visible five minutes. Across the eligible limpets, mean distance closure was 0.378 cm in the predator-cue-absent treatment and 0.878 cm in the predator-cue-present treatment (Fig.~\ref{fig:riding}B). The treatment difference, $\Delta D$, was +0.500 cm, with a chamber-cluster bootstrap 10--90\% interval of +0.188 to +0.814 cm.

Paired-trajectory decomposition captured projected movement by both hosts and limpets toward one another among eventual riders (Fig.~\ref{fig:riding}C). The host-snail component generally indicated movement toward the limpet, while some limpets made comparatively large projected movements toward the host, producing a right-skewed distribution of size-normalised limpet contributions. Among the 58 eventual riders in the nominal ten-minute analysis window (18 predator-cue-absent and 40 predator-cue-present), the limpet's size-normalised projected contribution exceeded the host's in 22 pairs (8 of 18 and 14 of 40, respectively). The approach pattern thus included contributions by both animals, with the relative contribution varying among pairs. Figure S1 in the electronic supplementary material provides a descriptive overview of all visible host-snail and limpet trajectories by chamber.

In the complementary mucus-exposure assay, the change in cumulative path length relative to each limpet's baseline varied less among individuals in the mucus groups than in the no-mucus group. We compared path lengths at 200 min between each limpet's assigned-surface run and its no-mucus baseline run. Paired observations were available for 22 limpets (seven assigned to \textit{Tegula} mucus, seven to \textit{Monodonta} mucus, and eight to no mucus). The observed standard deviations of $\log R_i$ were 1.054, 0.263, and 0.296 for no mucus, \textit{Tegula}, and \textit{Monodonta}, respectively. In a treatment-specific model, the posterior probabilities that the log-ratio standard deviation was lower than in the no-mucus group were 0.997 for \textit{Tegula} and 0.997 for \textit{Monodonta}. For no mucus, \textit{Tegula}, and \textit{Monodonta}, the estimated population medians of $R_i$ were 1.005, 1.139, and 0.730, respectively (posterior medians of $\exp(\mu_g)$). Relative to no mucus, the ratios of group medians were 1.13 (95\% posterior interval 0.57--2.27) for \textit{Tegula} and 0.73 (0.38--1.46) for \textit{Monodonta}. The largest no-mucus ratio was 6.940 (ID 03); the source video and tracked coordinates agreed at six sampled frames spanning the analysed interval. In a sensitivity analysis excluding this individual, the posterior probabilities of lower dispersion relative to no mucus were 0.982 for \textit{Tegula} and 0.971 for \textit{Monodonta}. The paired observations and model summaries are shown in Fig. S2. Together, these results show lower between-individual variation in the log-transformed paired ratios in the mucus groups.

\begin{figure}[htbp]
\centering
\includegraphics[width=138mm]{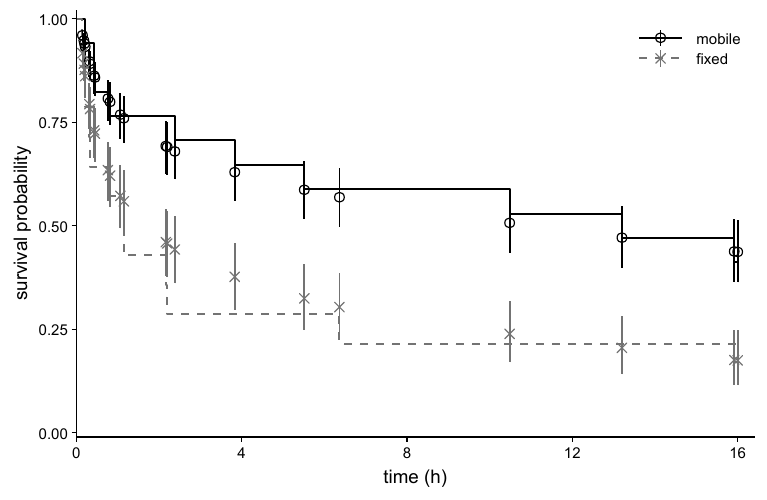}
\caption{Host mobility increased limpet survival probability. Black solid lines and open circles denote mobile hosts; grey dashed lines and crosses denote fixed hosts. Step lines show empirical survival, points show posterior median survival probabilities, and vertical bars show 25--75\% posterior intervals from the discrete-time hazard model. The posterior median hazard ratio for fixed versus mobile hosts was 2.111, and posterior survival at the end of the assay was higher for mobile hosts.}\label{fig:survival}
\end{figure}

Host mobility improved limpet survival under predation risk. This separate assay compared mobile and fixed living hosts, retaining shell occupancy in both treatments and testing a consequence of the host--rider association. In the mobile-host treatment, 10 of 17 trials ended in predation and 7 were right-censored; in the fixed-host treatment, 11 of 14 trials ended in predation and 3 were right-censored. The discrete-time log-link hazard model gave a posterior median hazard ratio of 2.111 for fixed versus mobile hosts (10--90\% posterior interval: 1.213--3.681). By the final observation time of 960 min, posterior median survival probability was 0.437 on mobile hosts but 0.175 on fixed hosts (Fig.~\ref{fig:survival}).

\FloatBarrier
\section{Discussion}
Our study takes the host's previously characterised critical-like locomotion as its starting point and shifts the focus to the rider: how it contributes to forming an interspecific association and what it gains from host-generated movement. The experiments connect this behavioural formation of coupling to a survival benefit from host mobility. Predator cues increased boarding and visible distance closure, and the trajectory decomposition revealed active contributions by limpets to movement toward the host within the observed analysis window. The mobile--fixed comparison demonstrated higher survival on mobile hosts even though limpets in both treatments occupied a host shell. In \textit{L. tenuisculpta}, already recognised as a facultative epizoite with broad host use \cite{nakayama2022}, these findings identify both a condition that promotes host use and a functional consequence of coupling to a moving host.

The formation of this association involved movement by both the rider and its host. Predator cues more than doubled the raw proportion of riders, from 19 of 80 to 42 of 76 limpets, and strengthened distance closure during the final visible five minutes. The movement decomposition addressed a complementary question: how much each animal contributed projected movement toward the other within the nominal ten-minute window among eventual riders. In 22 of the 58 pairs with valid trajectories, the limpet's size-normalised projected contribution exceeded the host's. These individuals exemplify active riders: small animals making a substantial normalised contribution to approaching a mobile host. Host movement also contributed to the observed approach, making the encounter an interaction between two moving animals.

The survival experiment supports a benefit of host movement once the association is formed. Predation hazard was higher on fixed hosts among the 31 retained observations, with a posterior median hazard ratio of 2.111 for fixed versus mobile hosts (10--90\% posterior interval: 1.213--3.681). Both treatments retained the shell of a living host as the limpet's substrate, supporting a survival benefit from host mobility beyond shell occupancy alone. This comparison should be interpreted in light of the nine exclusions among 40 observations after treatment assignment, including six fixed and three mobile, because differential exclusion could affect the result. Earlier work reported reduced predation in host-associated epizoic limpets \cite{mapstone1984,vermeij1987}. In particular, the \textit{Patelloida mufria} experiments compared living, active hosts with immobilised empty shells \cite{mapstone1984}. By using living hosts in both treatments, our experiment identifies host mobility as a contributor to protection in \textit{L. tenuisculpta}. Together with the approach analyses and earlier studies of epibiosis and gastropod hitchhiking \cite{wahl2008,kano2009,mapstone1984,vermeij1987}, this result connects the behavioural formation of the association to a host that serves as both a substrate and a beneficial source of movement.

The host's dynamical character provides the setting for this coupling. The spontaneous locomotion of \textit{T. nigerrima} has previously been characterised by statistical signatures consistent with near-critical dynamics over limited ranges \cite{kagaya2025}. A prospective rider must approach this moving target before gaining access to its locomotion. On the rider's side, predator cues increased riding and visible distance closure. The spectral analyses were compatible with altered movement, but the overall amplitude contrast and high-frequency slope contrast remained uncertain. The fixed-length sensitivity analysis gave a smaller amplitude contrast (electronic supplementary material). These results do not establish restlessness or a mechanism linking movement to boarding. Event-aligned measurements would help determine whether temporal features of limpet movement are related to the formation of the association.

Once the association forms, the host changes from an independently moving target into a source of locomotion for the rider. The survival assay supports the functional value of access to this movement. This relationship extends the question of adaptive dynamics from an individual animal to its association with another organism. Criticality has been proposed to support a large dynamical repertoire, sensitivity to inputs, and a balance between robustness and flexibility in neural and other living systems \cite{chialvo2010,munoz2018,shew2013}. Host riding adds an ecological possibility to this perspective: movement relevant to an animal's survival can be supplied partly by another animal. In this sense, the rider partly outsources useful locomotion through a behaviourally formed association.

Gastropod mucus is a candidate sensory route into this coupling. In the complementary assay, $\log R_i$ varied less among individuals in the mucus groups than in the no-mucus group. This result concerns between-individual variation in the paired response of solitary limpets. Mucus trails can support homing, aggregation, and mate searching and can mediate predator--prey interactions \cite{wada2026,ng2013}, providing a persistent local trace of recent movement. Such traces could guide a limpet's search through local responses. Establishing this route into host--rider coupling requires direct tests of directional attraction, trail following, boarding, and interactions with predator cues. The present assay used solitary limpets and the mucus donor \textit{T. rustica}, a proxy for the host species \textit{T. nigerrima}. Removing or masking fresh \textit{T. nigerrima} mucus while controlling waterborne host odour, or presenting artificial trails without a host, would test specific steps in the proposed route. Combining those manipulations with predator cues would test whether risk changes the use of mucus-linked information during approach.

The next step is to identify how rider and host dynamics contribute to the formation and survival value of this association. For the rider, aligning locomotor measurements with independently annotated approach and boarding events would test whether spectral slopes or other temporal features predict boarding and establish the temporal sequence needed to examine causal mechanisms. Dedicated dynamical measurements could test criticality in the limpet or matching between rider and host; the present connection to criticality rests on the prior characterisation of host locomotion rather than measurements of neural dynamics in either species. For the host, controlled motion playback, robotic hosts, or mechanically translated shells could test whether the temporal organisation of movement contributes to the survival benefit supported by the mobile--fixed comparison. Higher-resolution tracking could also resolve fine-scale movement on the host shell after boarding, extending the present decomposition of visible approach trajectories. Together, these experiments would clarify the mechanisms underlying an association whose behavioural formation and survival benefit are documented here.

Taken together, these findings provide a case study of how a small animal forms a survival-enhancing association with a mobile host whose exploratory locomotion has previously been characterised as critical-like \cite{kagaya2025}. Predator cues promote entry into this relationship, limpets contribute to approach, and host mobility provides a benefit beyond shell occupancy alone. The central advance is the connection between the rider's contribution to forming an interspecific association and the survival value of gaining access to host-generated movement.

\section{Ethics}
The authors determined that formal animal-ethics approval was not required because the gastropods and crab used here were invertebrates outside the animal-ethics review requirements applicable to this study. The authors also determined that collection permits were not required because the study species are common, non-protected intertidal taxa. Animals were collected by hand from intertidal habitat in Shirahama and Hiranai, Japan, maintained in flowing seawater before the assays, and handled only as required for the experiments. The procedures were designed to avoid unnecessary suffering and to use the minimum number of animals needed to address the study questions.

\section{Data accessibility}
Processed datasets and analysis scripts for the 2019 riding and survival assays are available from figshare \cite{dataset2026}. The corrected speed-derived files and revised spectrum analysis reported here, the tracking-curation audit, and the code and analysis files for the paired-ratio mucus analysis, are included in the updated release. Supplementary text and figures follow the main text in the combined preprint PDF. Supplementary Movie S1 is supplied separately as an all-chamber trajectory GIF animation; a static visualisation of these trajectories is also available from figshare \cite{dataset2026}.

\section{Declaration of AI use}
OpenAI ChatGPT and Codex were used during manuscript and analysis revisions. All manuscript sections, figure legends, electronic supplementary material, analysis code, and workflow documentation were edited or otherwise influenced by these tools. The authors defined the research questions, collected the data, constructed the initial statistical models, supplied manuscript drafts and outlines, and determined the scientific interpretations. The tools were used iteratively to revise and debug author-initiated analysis code, execute analyses, check outputs, rewrite text from author-provided drafts and outlines, and assist with language editing and proofreading. The authors retain responsibility for reviewing and verifying all code, analyses, numerical results, scientific content, interpretations, and final wording. They take full responsibility for the work.

\section{Authors' contributions}
R.N. contributed to experimental design, construction of the experimental setup, data collection, data analysis, and manuscript writing. R.N., T.N., and K.H. contributed to study planning, manuscript development, and the natural-history interpretation of the system. K.H. contributed to construction of the experimental setup and preparation of the Methods section. K.K. contributed to the overall intellectual framing of the study, including study design, analysis, interpretation, and manuscript development, but did not directly collect data. 

\section{Competing interests}
The authors declare no competing interests.

\section{Funding}
We acknowledge support from the Hakubi Project, Kyoto University.

\section{Acknowledgements}
We thank the staff of the Shirahama Aquarium, Kyoto University, for support with field logistics, specimen handling, and animal care. We are grateful to the Fisheries Research Institute, Aomori Prefectural Industrial Technology Research Center, for access to facilities and experimental equipment.

\end{document}


\maketitle

\noindent\textsuperscript{*}Ryo Nakayama and Katsushi Kagaya contributed equally to this work.\\Corresponding author: Katsushi Kagaya (\texttt{kkagaya@mail.kitami-it.ac.jp}).

\section{Overview}

This supplementary material describes the experimental setup, models, data processing, and complementary and sensitivity analyses for the accompanying manuscript. The main text focuses on predator cues, visible approach, movement toward the other animal during final approach, and the survival value of mobile hosts.

The rider species was the epizoic limpet \textit{Lottia tenuisculpta}; the host species was \textit{Tegula nigerrima}; and the predator-cue source was the carnivorous crab \textit{Leptodius affinis}. Coordinates used for locomotion and paired-trajectory analyses were obtained from time-lapse images by markerless tracking with DeepLabCut version 3.0.0 rc10 \cite{mathis2018}. Processed data and analysis scripts for the 2019 riding and survival assays are available from figshare \cite{dataset2026}. The corrected speed-derived files and revised spectrum analysis reported here, the tracking-curation audit, and the code and analysis files for the paired-ratio mucus analysis described in Section~\ref{sec:mucus}, are included in the updated release.

\section{Experimental setup}

For the riding and survival assays, \textit{L. tenuisculpta}, \textit{T. nigerrima}, and \textit{L. affinis} were collected from the nearshore area around the Seto Marine Biological Laboratory, Shirahama, Wakayama, Japan, between August and September 2019. Experimental limpets were collected exclusively from rock surfaces. Host snails were maintained separately in flowing seawater and cleared of any attached limpets before experiments. Crabs were individually identified, measured for carapace width, matched by size for paired trials, and held without food for two weeks before use. All limpets used in these assays were smaller than 7 mm in shell length (2.35--5.29 mm). Experimental individuals, cages, and positions were selected using dice-generated random numbers.

The riding and survival assays were conducted at the Shirahama Aquarium adjacent to the Seto Marine Biological Laboratory on 17--21 and 26--30 September and 1 October 2019. Trials ran overnight under illuminated conditions and were recorded from above at 20-s intervals with a GoPro HERO7 Black. Multiple chambers were recorded within one field of view. Paired-trajectory analyses used frames in which both host and limpet coordinates were available, because host shells could occlude limpets.

In the riding assay, each chamber contained one host snail and four individually marked limpets. Main-text Figure 1 shows the experimental layouts and overhead tracking setup. Water flow was started before each trial, the four limpets were placed at randomly selected starting positions on the chamber floor, and the host was placed at a fixed starting position. In predator-cue-present chambers, a crab was held in a separate compartment behind a perforated barrier, allowing waterborne cues to enter the test compartment while preventing physical contact; the corresponding compartment was empty in predator-cue-absent chambers. Twenty overnight trial sets, comprising 40 chambers, were initiated. One chamber was excluded because complete riding annotation was unavailable, leaving 39 chambers and 156 limpets.

The tracking dataset comprised 195 individual trajectories: 156 limpets and 39 host snails from the 39 retained chambers. Full chamber records contained 2,867--2,885 sampling times at 20-s intervals, spanning 15.92--16.02 h. Two limpet trajectories were shorter, containing 602 and 810 records; the other 193 trajectories contained 2,867--2,885 records each. Across all trajectories, the analysis input contained 557,591 animal--frame records, of which 430,826 had non-missing $x$--$y$ coordinates. These repeated records describe the temporal coverage of tracking, rather than independent biological replicates. The individual and chamber sample sizes, available-coordinate requirements, and observation windows for each analysis are specified below.

Tracking outputs from DeepLabCut were manually curated using custom tools that display the tracked position on the corresponding video frame and allow frame-by-frame navigation. Manual curation included both coordinate adjustment and the assignment of missing values to excluded positions. Subsequent analyses used these curated trajectories.

To document the extent of manual curation, we compared archived DeepLabCut H5 predictions with the corrected CSV coordinates after accounting for integer-pixel conversion and CSV numerical precision. For the 194 trajectories whose archived CSVs matched the current analysis input (554,708 animal--frame records), 5,322 positions had adjusted coordinates and 124,299 had been marked as missing. Of the missing positions, 115,261 lay in contiguous terminal sections after the last retained position. The remaining trajectory (ID 9-4) was retained in the analysis input but omitted from this curation tally because the correspondence between its archived H5 predictions and current analysis input remains unresolved. These totals count final changed positions, rather than editing operations or frames inspected.

The 40 crab records in the survival assay had distinct individual identifiers. Of the 20 crabs used in predator-cue-present chambers in the riding assay, 18 had previously been used in the survival assay and were reused in the riding assay one day later (16 crabs) or five days later (two crabs).

The survival assay compared limpets riding on mobile hosts with those riding on fixed hosts. Each observation involved one limpet on a living \textit{T. nigerrima} host and one crab in a test chamber. Each numbered trial paired one mobile-host observation with one fixed-host observation in separate chambers, with a different crab in each chamber. A limpet was placed on the aperture-facing side of the first body whorl of each host shell; the limpet itself was not restrained. A small amount of instant adhesive was applied to the underside of the host in both treatments. In the fixed-host treatment, the host was secured to the chamber floor. In the mobile-host treatment, the adhesive dried without the host being secured to any surface. After approximately 10 min of drying, seawater was added and a crab was introduced at the start of observation. Time to predation was read from the video. Forty observations in 20 treatment pairs were initiated. Four fixed hosts lost their attachment to the chamber, and five limpets detached from their hosts (two fixed-host and three mobile-host trials). These nine observations lacked a valid predation endpoint and were excluded entirely. Of the remaining 31 observations, 17 were mobile-host and 14 fixed-host trials. Limpets that remained on their hosts and were alive at the end of observation were right-censored (seven mobile and three fixed). The primary analysis retained valid members of otherwise incomplete treatment pairs.

For the complementary mucus-exposure assay, \textit{L. tenuisculpta} and the mucus-donor snails \textit{Tegula rustica} and \textit{Monodonta confusa} were collected at Hiranai, Aomori, Japan, on 11 July 2025 and maintained separately in flowing seawater. \textit{T. rustica}, another species in the same genus as the host \textit{T. nigerrima}, was used as a mucus-donor proxy; \textit{M. confusa} was included as an abundant sympatric gastropod rarely used as a host by \textit{L. tenuisculpta}. Four overnight runs began on 16, 17, 22, and 23 July 2025 and were recorded every 5 s with a GoPro HERO8 Black. The same identified limpets were observed first on no-mucus surfaces and then in dishes assigned to no mucus, \textit{Tegula} mucus, or \textit{Monodonta} mucus.

\section{Model and analysis details}

\subsection{Host-riding probability}

We tested whether the probability of riding at least once differed between predator-cue treatments, while accounting for the grouping of four limpets within each chamber.

Let $Y_i=1$ if limpet $i$ rode on the host at least once during the overnight observation period and $Y_i=0$ otherwise. Let $x_i=1$ for predator cue present and $x_i=0$ for predator cue absent, and let $c(i)$ identify the chamber containing limpet $i$. The fitted model was
\begin{align}
Y_i &\sim \operatorname{Bernoulli}(p_i), \label{eq:riding-response}\\
\log\frac{p_i}{1-p_i} &= \alpha+\beta x_i+b_{c(i)},
& b_c &\sim \operatorname{Normal}(0,\sigma_{\mathrm{chamber}}). \label{eq:riding-model}
\end{align}
The chamber random intercept $b_c$ accounts for the grouping of four limpets within each chamber. Thus, individuals sharing a chamber are not treated as independent replicates of the predator cue treatment. The intercept $\alpha$ gives the predator-cue-absent log odds at $b_c=0$, and $\beta$ is the predator-cue-associated change in log odds. The reported riding probabilities are $\operatorname{logit}^{-1}(\alpha)$ and $\operatorname{logit}^{-1}(\alpha+\beta)$, with the chamber effect set to zero; the predator cue odds ratio is $\exp(\beta)$. Priors were $\alpha,\beta\sim\operatorname{Normal}(0,2.5)$ and $\sigma_{\mathrm{chamber}}\sim\operatorname{Exponential}(1)$. Normal distributions here and below use a mean and standard deviation; exponential distributions use a rate. Chamber effects were implemented in non-centred form, $b_c=\sigma_{\mathrm{chamber}}z_c$ with $z_c\sim\operatorname{Normal}(0,1)$, following multilevel-modelling principles \cite{gelman2007}.

The final dataset comprised 156 limpets from 39 chambers: 80 predator-cue-absent and 76 predator-cue-present individuals. The model used four chains of 4,000 iterations, including 1,000 warmup iterations (seed 1234).

\subsection{Paired trajectories and movement components}\label{sec:paired-trajectories}

We addressed two distinct questions: whether closure of the visible distance between host and limpet differed between predator-cue treatments, and how each animal contributed projected movement toward the other among eventual riders. The distance-closure analysis included all eligible limpets, whereas the movement-component analysis was restricted to eventual riders and used a separately defined reference frame.

For paired trajectories, we matched host and limpet coordinates from the same chamber and image frame, retaining only frames in which both animals were visible. Host--limpet distance was converted from pixels to centimetres using 10 cm / 500 px = 0.02 cm per pixel. The final visible paired frame for each limpet was aligned to time 0; this reference frame does not necessarily coincide with the moment of boarding. Distance closure over the final visible five minutes was calculated as the median distance in the first three samples of the window minus the median distance in the final three samples. For $\Delta D$, the difference in mean distance closure between treatments (cue present minus cue absent), we estimated uncertainty using 5,000 bootstrap replicates (seed 42). Within each predator cue treatment, chambers were sampled with replacement, and all eligible limpets from each selected chamber were retained; the 10th and 90th percentiles of the replicate differences were reported. The analysis included 148 limpets with at least six visible paired frames in the final five-minute window (74 per predator cue treatment, across 39 chambers).

For eventual riders, we quantified each animal's projected movement toward the other. This decomposition used the movement-analysis reference frame stored as \path{riding_frame} in \path{data/inputs/merged_limept.csv}, a different endpoint from that used in the distance-closure analysis. This frame is the final complete limpet-track record after records with missing values were removed; both coordinates and speed were required. Paired coordinates after this frame were excluded, and the frame was aligned to time 0. Four eventual riders had later visible coordinate pairs but missing limpet speed, so their movement-analysis reference frame preceded the last paired-coordinate frame. The stored field defines the endpoint of the movement-analysis window rather than an independently annotated boarding time. The same endpoint determines the eventual-rider positions in main-text Figure 2A, whereas the binary riding model uses whether each limpet rode during the trial.

The final ten-minute decomposition required at least 15 valid 20-s movement segments formed from consecutive paired frames. Valid segments also required a positive initial host--limpet separation and displacement no greater than 25 pixels for the limpet or 100 pixels for the host. Of the 61 riders, IDs 21-2 and 35-3 in the predator-cue-present treatment had only 3 and 4 valid segments, respectively, and ID 40-2 in the predator-cue-absent treatment had only 4. These three were excluded from the ten-minute decomposition, leaving 58 riders (40 predator-cue-present and 18 predator-cue-absent).

The main-text decomposition used the ten-minute window ending at this reference frame. The five-minute window ending at the same frame provided a shorter-window check. Each 20-s segment was selected by the time of its later frame, including endpoints from $-10$ to 0 min (or $-5$ to 0 min in the short-window check). Thus a segment ending exactly at the window start can begin 20 s earlier; the displayed cumulative sums and component totals use this endpoint-inclusive convention.

For each 20-s step, the host--limpet axis at the previous frame defined the reference direction. Limpet movement toward the host and host-snail movement toward the limpet were summed separately; positive values indicate projected displacement toward the other animal along the axis defined at the start of each step. The sum of the two projections is not an exact decomposition of the finite-step change in Euclidean separation. For the body-size-normalised display, limpet movement toward the host was divided by limpet shell length, and host-snail movement toward the limpet was divided by host shell width.

\subsection{\texorpdfstring{Skew-$t$}{Skew-t} model for own-size-normalised movement}

We used a descriptive Bayesian skew-$t$ model to summarise the asymmetric, heavy-tailed distributions of own-size-normalised movement components. The model compared group location parameters across animal types and predator-cue treatments while accounting for paired components from the same rider--host pair.

We fitted the model separately to the five- and ten-minute windows. Let $y_{ia}$ be the component for animal $a$ (limpet or host snail) in rider pair $i$, and let $g(i,a)$ denote its group, defined by predator-cue treatment and animal type. There are four groups: predator-cue-absent limpet, predator-cue-absent host, predator-cue-present limpet, and predator-cue-present host. The location for each observation was
\begin{equation}
\xi_{ia}=\mu_{g(i,a)}+b_i,
\qquad b_i\sim\operatorname{Normal}(0,\tau).
\label{eq:skew-location}
\end{equation}
The same rider-level intercept $b_i$ enters the limpet and host components of a pair. With $z=(y-\xi)/\omega$, the density implemented in Stan was
\begin{equation}
f(y\mid\xi,\omega,\alpha,\nu)
=\frac{2}{\omega}\,t_\nu(z)\,
T_{\nu+1}\!\left(\alpha z\sqrt{\frac{\nu+1}{\nu+z^2}}\right),
\qquad \omega>0,\quad\nu>2.
\label{eq:skew-density}
\end{equation}
Here $t_\nu$ is the standard Student-$t$ density and $T_\nu$ its cumulative distribution function. Each group defined by predator-cue treatment and animal type had its own location $\mu_g$, residual scale $\omega_g$, and skewness $\alpha_g$, whereas all groups shared the degrees of freedom $\nu$. The priors were $\mu_g\sim\operatorname{Normal}(0,4)$, $\omega_g\sim\operatorname{Lognormal}(0,1)$, $\alpha_g\sim\operatorname{Normal}(0,4)$, $\tau\sim\operatorname{Exponential}(1)$, and $\nu-2\sim\operatorname{Exponential}(0.1)$. The lognormal parameters are the mean and standard deviation on the natural-log scale. Rider effects were implemented as $b_i=\tau z_i$ with $z_i\sim\operatorname{Normal}(0,1)$. Contrasts between the $\mu_g$ values summarise differences in model location. They are distinct from differences in observed medians and do not necessarily represent differences in distribution means. Each fit used four chains of 4,000 iterations, including 1,000 warmup iterations (seed 20260629).

\subsection{Discrete-time predation hazard}

This analysis tested whether predation risk differed between limpets on mobile and fixed living hosts, with shell occupancy retained in both treatments.

We analysed time-to-predation data using a discrete-time hazard model with a log link \cite{singer1993}. Let $x_i=0$ for a mobile host and $x_i=1$ for a fixed host. The ordered distinct observation times, in minutes, were $t_1,\ldots,t_K$, with $t_0=0$ and $\Delta_k=t_k-t_{k-1}$. The conditional probability of predation in one minute was constant within each interval $(t_{k-1},t_k]$ and was specified by
\begin{equation}
\log h_{ik}=a_k+\beta x_i,
\qquad h_{ik}=\exp(a_k+\beta x_i).
\label{eq:survival-hazard}
\end{equation}
Thus $a_k$ is the mobile-host baseline log hazard and $\exp(\beta)$ is the fixed-to-mobile ratio of the one-minute hazard probabilities. The probability of surviving through time $t_k$ is
\begin{equation}
S_i(t_k)=\prod_{\ell=1}^{k}(1-h_{i\ell})^{\Delta_\ell}.
\label{eq:survival-probability}
\end{equation}
An observed predation at $t_k$ contributes $S_i(t_{k-1})(1-h_{ik})^{\Delta_k-1}h_{ik}$ to the likelihood, whereas right-censoring at $t_k$ contributes $S_i(t_k)$; $S_i(t_0)=1$. The baseline log hazard had a first-order random-walk prior: $a_1\sim t_3(0,5)$ and $a_k\sim\operatorname{Normal}(a_{k-1},\sigma_a)$ for $k\geq2$, with $\sigma_a\sim\operatorname{half\mbox{-}Cauchy}(0,1)$. The treatment coefficient had prior $\beta\sim t_3(0,2.5)$, where the Student-$t$ parameters denote degrees of freedom, location and scale. Baseline log hazards were constrained to be at most $-10^{-6}$ in Stan, and both treatment hazards must remain below 1 for a valid likelihood. The primary model used all 31 valid observations (17 mobile and 14 fixed), retained right-censored observations, and did not include the trial-pair identifier. It used four chains of 5,000 iterations, including 500 warmup iterations (seed 3).

The primary hazard ratio describes the 31 retained observations, including right-censored outcomes. The nine exclusions occurred after assignment and were more frequent in the fixed-host treatment (six versus three). The hazard ratio therefore describes a retained-sample contrast, not an intention-to-treat estimate for all 40 initiated observations.

The nine excluded observations were: fixed-host loss of attachment to the chamber in trials 2, 10, 12, and 20; limpet detachment from the fixed host in trials 6 and 19; and limpet detachment from the mobile host in trials 7, 14, and 17. The cause of each loss of host fixation was not identified. The distinction between these exclusions and end-of-observation right-censoring was maintained in the primary analysis.

The host-riding, movement-component, predation-hazard, locomotor-spectrum, and paired mucus-exposure models were fitted in Stan \cite{carpenter2017,stan238}. The locomotor-spectrum model, sampling diagnostics, and independent numerical-integration check are described in Section~\ref{sec:spectrum}; the paired mucus-exposure model and its diagnostics are specified in Section~\ref{sec:mucus}.

\section{Trajectory visualisation files}

The trajectory map and animation provide a descriptive overview of host-snail and limpet paths across chambers and predator-cue treatments, placing the selected paired-trajectory windows in the context of the recorded movements.

The static all-chamber trajectory map is shown as Supplementary Figure S1. A version of this map with different colours and treatment labels is available from figshare as the files \path{all_snail_limpet_trajectories_by_chamber.pdf} and \path{all_snail_limpet_trajectories_by_chamber.png}. One all-chamber GIF animation accompanies this map as Supplementary Movie S1. These visualisations use the same processed DeepLabCut-derived coordinate table as the paired-trajectory analyses. They were used descriptively, rather than as additional inferential tests, to place the distance and movement summaries in the context of the recorded paths. The GIF samples the overnight time-lapse sequence into 100 animation frames at 10 frames s\(^{-1}\), with fading trails showing recent movement history.

\section{Projected movement toward the other animal among eventual riders}

We quantified each animal's projected movement toward the other to describe observed approach among eventual riders. The main-text display uses the nominal ten-minute window ending at each rider's movement-analysis reference frame (Section~\ref{sec:paired-trajectories}). It includes riders with valid paired trajectories: 18 predator-cue-absent and 40 predator-cue-present riders. We also report the five-minute window ending at the same reference frame to assess how these descriptive summaries depend on the length of the analysis window.

In the final ten-minute window, median limpet movement toward the host was -0.006 cm in predator-cue-absent riders and 0.168 cm in predator-cue-present riders. Median host-snail movement toward the limpet was 1.54 cm and 1.33 cm, respectively. Host-snail movement toward the limpet was larger than limpet movement toward the host in 11 of 18 predator-cue-absent riders and 26 of 40 predator-cue-present riders.

We also expressed each component relative to the animal's own body size, because the limpet is much smaller than the host snail. Limpet movement toward the host was normalised by limpet shell length, and host-snail movement toward the limpet by host shell width. In the final ten-minute window, median limpet movement toward the host was -0.018 shell lengths in predator-cue-absent riders and 0.437 shell lengths in predator-cue-present riders, whereas median host-snail movement toward the limpet was 0.944 and 0.751 shell widths. The limpet component had a broad upper tail: its 90th percentiles were 9.55 and 8.15 shell lengths in predator-cue-absent and predator-cue-present riders, respectively. Limpet movement toward the host exceeded one limpet shell length in 7 of 18 predator-cue-absent riders and 16 of 40 predator-cue-present riders; host-snail movement toward the limpet exceeded one host shell width in 9 of 18 and 18 of 40 riders.

In the final ten-minute window, the own-size-normalised limpet component exceeded that of its paired host snail in 8 of 18 predator-cue-absent riders and 14 of 40 predator-cue-present riders (22 of 58 overall). This comparison concerns the relative components within each pair, whereas the preceding counts concern the one-shell-length threshold and the check below uses a five-minute window. The ten-minute trajectories and component values are shown in Supplementary Figure S3.

The final five-minute window provided a shorter-window check on the same pattern. In absolute distance, median limpet movement toward the host was 0.150 cm in predator-cue-absent riders and 0.198 cm in predator-cue-present riders. Median host-snail movement toward the limpet was 1.708 cm and 1.742 cm, respectively. On the own-size scale, median limpet movement toward the host was 0.351 shell lengths in predator-cue-absent riders and 0.519 shell lengths in predator-cue-present riders, whereas median host-snail movement toward the limpet was 1.052 and 0.966 shell widths. Limpet movement toward the host exceeded one limpet shell length in 8 of 18 predator-cue-absent riders and 14 of 40 predator-cue-present riders; host-snail movement toward the limpet exceeded one host shell width in 9 of 18 and 20 of 40 riders. These five-minute summaries are a shorter-window check, not the ten-minute values displayed in Supplementary Figure S3.

The rider-level skew-$t$ model summarises group location parameters while allowing asymmetric distributions. We fitted the same model to the own-size-normalised five- and ten-minute values; its location summaries are distinct from direct medians of the observed distributions. In the five-minute model, the posterior median snail-minus-limpet location contrast was 0.497 in the predator-cue-absent treatment (10--90\% interval: 0.139 to 1.039) and 0.363 in the predator-cue-present treatment (0.147 to 0.640). Predator cue effects within each animal type were weaker: the posterior median predator cue effect was 0.124 for limpet movement (-0.244 to 0.557) and 0.001 for host-snail movement (-0.284 to 0.236).

In the ten-minute model, the posterior median snail-minus-limpet location contrast was 2.087 in the predator-cue-absent treatment (0.792 to 3.521) and 0.763 in the predator-cue-present treatment (0.305 to 1.341). The posterior median predator cue effect was 0.404 for limpet movement (-0.533 to 1.564) and -0.806 for host-snail movement (-1.900 to 0.046). Thus, the ten-minute analysis supports the same qualitative point that both animals can contribute projected movement toward one another, while also showing that the relative component estimates depend on the length of the visible approach window.

Both skew-$t$ fits sampled without divergent transitions or maximum tree-depth hits. The maximum \(\hat{R}\) was 1.002 for the five-minute model and 1.001 for the ten-minute model. The posterior median degrees-of-freedom parameter was 2.32 (10--90\% interval: 2.05 to 3.23) for the five-minute model and 2.49 (2.07 to 4.42) for the ten-minute model.

Posterior predictive checks showed that the model captured the broad asymmetry of the observed component distributions (Supplementary Figure S4). These checks assessed the shape and tails of the distributions, although observations in the long upper tails remained sparse.

\section{Locomotor spectrum}\label{sec:spectrum}

We examined predator-cue-associated variation in amplitude over the observed frequency range and, separately, in the high-frequency slope of limpet locomotion. Pooled spectra are descriptive; the amplitude comparison uses matched trial sets and a recording-date bootstrap, whereas the hierarchical slope model accounts for chamber grouping and slope-estimation uncertainty.

Frame-to-frame movement was calculated as the Euclidean displacement between consecutive frames within each individual's track. The first record of each track and steps with missing coordinates at either endpoint were assigned missing movement values. For each limpet, we retained observations from the longest available runs with complete speed data. When runs shared the maximum length, we processed their retained records together. Stored frame-to-frame movement values were converted from pixels to centimetres using 0.02 cm pixel$^{-1}$. The sampling frequency was 0.05 Hz (one sample every 20 s), while amplitudes retained units of cm frame$^{-1}$. We computed the discrete Fourier transform after subtracting the mean of the retained speed series. The amplitude at each frequency was the modulus of the transform divided by the number of samples. The reported quantity is therefore an amplitude spectrum; we used the modulus rather than the squared modulus that would define a power spectrum. No additional taper or linear detrending was applied in this Fourier-processing step. The implementation retained the first $\lfloor N/2\rfloor$ transform entries, including the zero-frequency entry, where $N$ is the number of samples. Zero-frequency and non-positive-amplitude values were then removed before logarithmic transformation; amplitudes were not doubled.

We constructed the descriptive spectrum by pooling spectral observations within each treatment into 16 equally spaced bins on the $\log_{10}$-frequency scale. The observed range was $-4.7610$ to $-1.6024$ in $\log_{10}$ Hz (approximately $1.73\times10^{-5}$ to $2.50\times10^{-2}$ Hz). Each point in main-text Figure 3A is the median $\log_{10}$ amplitude across pooled ordinates within a treatment and bin, at the bin midpoint. Longer records contribute more ordinates; the points are not equally weighted individual medians. Thin chamber curves and bands summarise individual medians in 64 finer bins. In Figure 3B, both groups are centred on the pooled cue-absent median for the corresponding 16-bin interval. Within each bin, the same kernel bandwidth and width scale are used for both groups, but the scale varies across bins. The densities use all ordinates and are descriptive. The displayed $-2$ to $+2$ range excludes the locations of 18 ordinates (17 below and one above) from view, not from calculation; the maximum fraction of any one displayed density shape outside this range was 0.116\%.

For a complementary treatment contrast, let $m_{icb}$ be the median log-amplitude of the available ordinates of individual $i$ in chamber $c$ and bin $b$. We averaged individual medians within each chamber and bin, then subtracted the cue-absent chamber mean from the cue-present chamber mean within each trial set. Within each bin, we averaged the paired differences over trial sets with observations from both treatments; the final contrast gives equal weight to the 16 bin-specific means. The experimental workbook records one chamber per treatment for each of 20 trial sets. Cross-checking the workbook against limpet identities and shell lengths, the existing mapping, and the documented correction for animal 25-1 established the trial-to-chamber correspondence. One cue-present chamber was excluded from the source data, leaving 19 complete pairs with 76 cue-absent and 75 cue-present limpets contributing positive-frequency spectra. The lowest-frequency bin had both treatments in only 10 trial sets from seven recording dates; the other bins had 14--19 complete sets. Each date contained two trial sets in the original record.

We sampled the ten recording dates with replacement, retaining all complete trial pairs, chambers, individuals and frequencies within each sampled date. The 10th--90th and 2.5th--97.5th percentiles of 20,000 date-bootstrap replicates (seed 20261107) define the 80\% and 95\% intervals. One replicate lacked a complete pair in at least one bin and was omitted from both intervals. The resulting 19,999 replicates are conditional on observed spectral availability and complete pairs. With only ten recording dates and sparse low-frequency observations, these percentile intervals are approximate; they are not simultaneous intervals for every frequency and do not resolve the mechanism of treatment allocation.

The equal-bin mean of within-trial differences was 0.082 log$_{10}$ amplitude units (80\% interval 0.014--0.156; 95\% interval $-0.015$ to $0.202$), corresponding to a ratio of geometric summaries of 1.21 (80\% interval 1.03--1.43; 95\% interval 0.97--1.59). The pooled cue-present treatment medians were higher in all 16 bins, whereas medians of individual bin medians were higher in 14 bins and the paired trial-mean contrast was positive in 15. These summaries have different weights. As an alternative using all 39 chambers and independently resampling chambers within treatment, without retaining trial or date pairing, the equal-bin contrast was 0.062 ($-0.034$ to $0.155$; 80\% interval), or a ratio of 1.15 (0.93--1.43). This is a differently weighted sensitivity comparison, not the paired-trial estimand.

All spectra used 20-s sampling, mean subtraction, no taper or additional linear detrending, and Fourier modulus divided by sample count. However, longest-run lengths differed: among the 155 animals contributing positive-frequency ordinates, median lengths were 1,359.5 frames without cues and 1,106 with cues. Length changes the Fourier grid and can affect normalised amplitude. As a sensitivity analysis we recomputed spectra using the first 512 consecutive valid movements from the selected run in each eligible individual. This retained 70 cue-absent and 57 cue-present animals in 20 and 18 chambers; 18 complete trial pairs contained 62 and 57 animals, respectively. The resulting Fourier grid occupied 12 of the original 16 log-frequency bins (bins 4 and 6--16), with frequencies $9.77\times10^{-5}$ to $2.49\times10^{-2}$ Hz. Using the paired estimator and date bootstrap on these 12 bins gave a difference of 0.018 (80\% interval $-0.103$ to $0.143$), or a ratio of 1.04 (0.79--1.39; 20,000 replicates, seed 20261108). This is a sensitivity check, not a direct estimate of the original full-range estimand: eligibility, retained time segment, frequency grid and bin coverage all differ.

Individual high-frequency slopes were estimated from unbinned spectral observations with log-frequency values of -3.1 or higher, rather than from the pooled display points. One predator-cue-present individual had no positive amplitude values in this range after fast Fourier transform preprocessing, and two additional individuals had fewer than three spectral points. The final chamber-level hierarchical normal model included 79 predator-cue-absent and 74 predator-cue-present individuals.

For individual $i$, an ordinary least-squares regression of $\log_{10}$ amplitude on $\log_{10}$ frequency in the selected high-frequency range yielded a slope estimate $\widehat q_i$ and its standard error $s_i$ (bounded below by $10^{-6}$ before fitting). The hierarchical model was
\begin{align}
\widehat q_i &\sim\operatorname{Normal}\!\left(\alpha+\beta x_i+b_{c(i)},\sqrt{s_i^2+\sigma_{\mathrm{individual}}^2}\right), \label{eq:spectrum-model}\\
b_c &\sim\operatorname{Normal}(0,\sigma_{\mathrm{chamber}}).
\end{align}
Here $x_i$ indicates predator cue presence and $c(i)$ is chamber identity. The supplied slope-estimation uncertainty $s_i$ and additional between-individual variability $\sigma_{\mathrm{individual}}$ enter the observation variance separately. Priors were $\alpha,\beta\sim\operatorname{Normal}(0,2.5)$ and $\sigma_{\mathrm{chamber}},\sigma_{\mathrm{individual}}\sim\operatorname{Exponential}(1)$. Chamber effects were implemented in non-centred form, $b_c=\sigma_{\mathrm{chamber}}z_c$ with $z_c\sim\operatorname{Normal}(0,1)$. The treatment summaries are $\alpha$ and $\alpha+\beta$, with chamber effects set to zero, and the predator cue contrast is $\beta$. This existing slope model includes chamber effects but no trial-set or recording-date effects, so its posterior interval does not explicitly account for those potential dependencies.

We fitted the model in Stan 2.32.2 through RStan 2.32.6, using R 4.4.2 and StanHeaders 2.32.10. Four chains each ran for 4,000 iterations, including 1,000 warmup iterations, yielding 12,000 retained posterior draws in total (seed 1234). The posterior summaries reported below are from these draws.

For the 43 sampled model parameters ($\alpha$, $\beta$, $\sigma_{\mathrm{chamber}}$, $\sigma_{\mathrm{individual}}$, and the 39 $z_c$ values), the maximum $\hat R$ was 1.001861, the minimum bulk effective sample size was 3,180.109, and the minimum tail effective sample size was 2,558.234. These summaries exclude transformed parameters and generated quantities. Sampling produced no divergent transitions and no hits at the maximum tree depth of 10; the largest observed tree depth was 5.

As an independent check, we analytically marginalised the Gaussian parameters $\alpha$, $\beta$, and the chamber effects, then integrated numerically over $\sigma_{\mathrm{chamber}}$ and $\sigma_{\mathrm{individual}}$. We used a two-dimensional midpoint grid with spacing 0.002 over the interval from 0 to 1.5 for each standard deviation. Quadrature-based posterior quantiles of $\alpha$, $\alpha+\beta$, and $\beta$ were obtained from the resulting mixtures of conditional normal distributions. Halving the grid spacing from 0.004 to 0.002 changed these quadrature quantiles by less than $2\times10^{-8}$. An independent Gauss--Legendre calculation, including an expanded integration domain from 0 to 3 for each standard deviation, confirmed the quadrature results. For $\alpha$, $\beta$, $\sigma_{\mathrm{chamber}}$, and $\sigma_{\mathrm{individual}}$, posterior means from Stan and numerical integration differed by less than 1.2 Monte Carlo standard errors of the corresponding Stan mean estimates.

The posterior median high-frequency slope was -0.445 in the predator-cue-absent treatment (10--90\% posterior interval: -0.475 to -0.414) and -0.428 in the predator-cue-present treatment (-0.459 to -0.397). The posterior median predator-cue-present minus predator-cue-absent contrast was 0.017 (10--90\% posterior interval: -0.027 to 0.060), with $P(\beta>0\mid\mathrm{data})=0.685$. The contrast remains uncertain, with the central 80\% posterior interval spanning zero. Whether high-frequency spectral structure relates to approach or boarding, and how it relates to living-host motion, remain questions for targeted mechanistic experiments.

\section{Complementary mucus-exposure assay}\label{sec:mucus}

This complementary, exploratory analysis compared cumulative locomotion after mucus assignment with each limpet's no-mucus baseline. We asked whether mucus assignment was associated with changes relative to baseline and whether the location and between-individual dispersion of these paired changes differed among treatments.

We analysed the paired observations at a common observed endpoint, 200 min after each run began. The first run for each identified limpet was on a no-mucus surface (baseline); in the later run, the same limpet was assigned to \textit{Tegula}, \textit{Monodonta}, or no mucus. We summed calibrated Euclidean displacements between consecutive processed coordinates through 200 min to obtain cumulative path length $D$ in centimetres. The paired response was $R_i=D_{i,\mathrm{assigned}}/D_{i,\mathrm{baseline}}$; it is dimensionless. The processed coordinate series is sampled at 10-s intervals. The 200-min endpoint was chosen after data collection, and the analysis is exploratory.

The 200-min analysis included 22 complete pairs: seven \textit{Tegula}, seven \textit{Monodonta}, and eight no mucus. The no-mucus reassignment group provides a repeated-measurement comparison. One \textit{Tegula}-assigned limpet (ID 13) left its dish during recording and was excluded from the assay. Of the remaining 23 assigned-run individuals, one \textit{Monodonta}-assigned limpet (ID 18) had no observed baseline coordinate at 200 min and could not contribute a paired response at that time. No individual was excluded from the primary 200-min analysis because of its response magnitude.

Let $g(i)$ denote the mucus treatment assigned to limpet $i$. The response and model were
\begin{align}
R_i &= \frac{D_{i,\mathrm{assigned}}}{D_{i,\mathrm{baseline}}},
& y_i &= \log R_i, \label{eq:mucus-response}\\
y_i &\sim \operatorname{Normal}\!\left(\mu_{g(i)},\sigma_{g(i)}\right),
& \mu_g &\sim \operatorname{Normal}(0,1), \label{eq:mucus-model}\\
\sigma_g &\sim \operatorname{half\mbox{-}Normal}(0,0.5).
\end{align}
Here $D$ is cumulative path length through 200 min, $R_i$ is a dimensionless within-individual ratio, and the logarithm is natural. The normal distribution is parameterised by its mean and standard deviation. The model estimates a separate mean $\mu_g$ and standard deviation $\sigma_g$ for each treatment, without pooling these parameters across treatments. Thus $\exp(\mu_g)$ is the population median of $R_i$ in treatment $g$, whereas $\sigma_g$ measures the dispersion of $\log R_i$ among individuals.

We fitted the model in the Stan program \path{mucus_paired_ratio.stan} using four No-U-Turn Sampler (NUTS) chains, 1,500 warmup and 2,000 retained iterations per chain, seed 20260928, \path{adapt_delta=0.95}, and \path{max_treedepth=12}. For the six treatment-level mean and standard-deviation parameters, the primary fit (half-normal scale 0.5) had a maximum $\hat R$ of 1.00057, a minimum bulk effective sample size of 5,647, and a minimum tail effective sample size of 4,230. With the wider half-normal scale 1.5, the corresponding values were 1.00093, 4,166, and 3,343. Neither fit had divergent transitions or maximum tree-depth hits.

At 200 min, the observed standard deviations of $\log R_i$ were 0.263 for \textit{Tegula}, 0.296 for \textit{Monodonta}, and 1.054 for no mucus. The posterior median $\sigma_g$ values were 0.292, 0.326, and 0.917, respectively. Under the primary prior, the posterior probabilities $P(\sigma_{\mathrm{Tegula}}<\sigma_{\mathrm{no\ mucus}})$ and $P(\sigma_{\mathrm{Monodonta}}<\sigma_{\mathrm{no\ mucus}})$ were 0.997 and 0.997, respectively. The posterior medians of $\exp(\mu_g)$ were 1.139, 0.730, and 1.005 in the same order. Relative to no mucus, the ratios of group medians were 1.13 (95\% posterior interval 0.57--2.27) for \textit{Tegula} and 0.73 (0.38--1.46) for \textit{Monodonta}. These intervals include 1.

For a new individual within each group, the 10th--90th percentile posterior predictive ranges for $R$ were 0.742--1.712 for \textit{Tegula}, 0.457--1.180 for \textit{Monodonta}, and 0.279--3.544 for no mucus. These central 80\% intervals describe a dimensionless paired ratio.

We assessed sensitivity to the prior on between-individual dispersion by repeating the fit with a wider half-normal(0,1.5) prior on $\sigma_g$. Under this wider prior, the two posterior probabilities of lower dispersion were 0.997 and 0.992; the no-mucus predictive range widened to 0.202--4.723.

A targeted data check examined the no-mucus individual with the largest observed response (ID 03, $R=6.940$). Its assigned-run source video and tracked coordinates agreed at six sampled frames spanning the 200-min interval. All 5,801 coordinates in the individual CSV matched the processed analysis table after truncation to integer pixels. This targeted check does not constitute frame-by-frame validation. To assess the influence of this largest observed no-mucus response, we excluded ID 03 only in a sensitivity fit, which gave corresponding probabilities 0.982 and 0.971 under the primary prior.

The exploratory comparison supports lower dispersion of $\log R_i$ among individuals in the two mucus groups than in the no-mucus group. This result concerns variation among individuals in a paired cumulative response, with causal interpretation constrained by small group sizes, the exploratory endpoint, and the baseline-first run order. Larger cohorts with varied run orders and direct measurements of within-run speed fluctuations, directional attraction, trail following, and boarding would help test whether this dispersion pattern relates to host finding.

\clearpage
\begin{landscape}
\thispagestyle{plain}
\begin{figure}[p]
\centering
\includegraphics[width=0.98\linewidth,height=0.83\textheight,keepaspectratio]{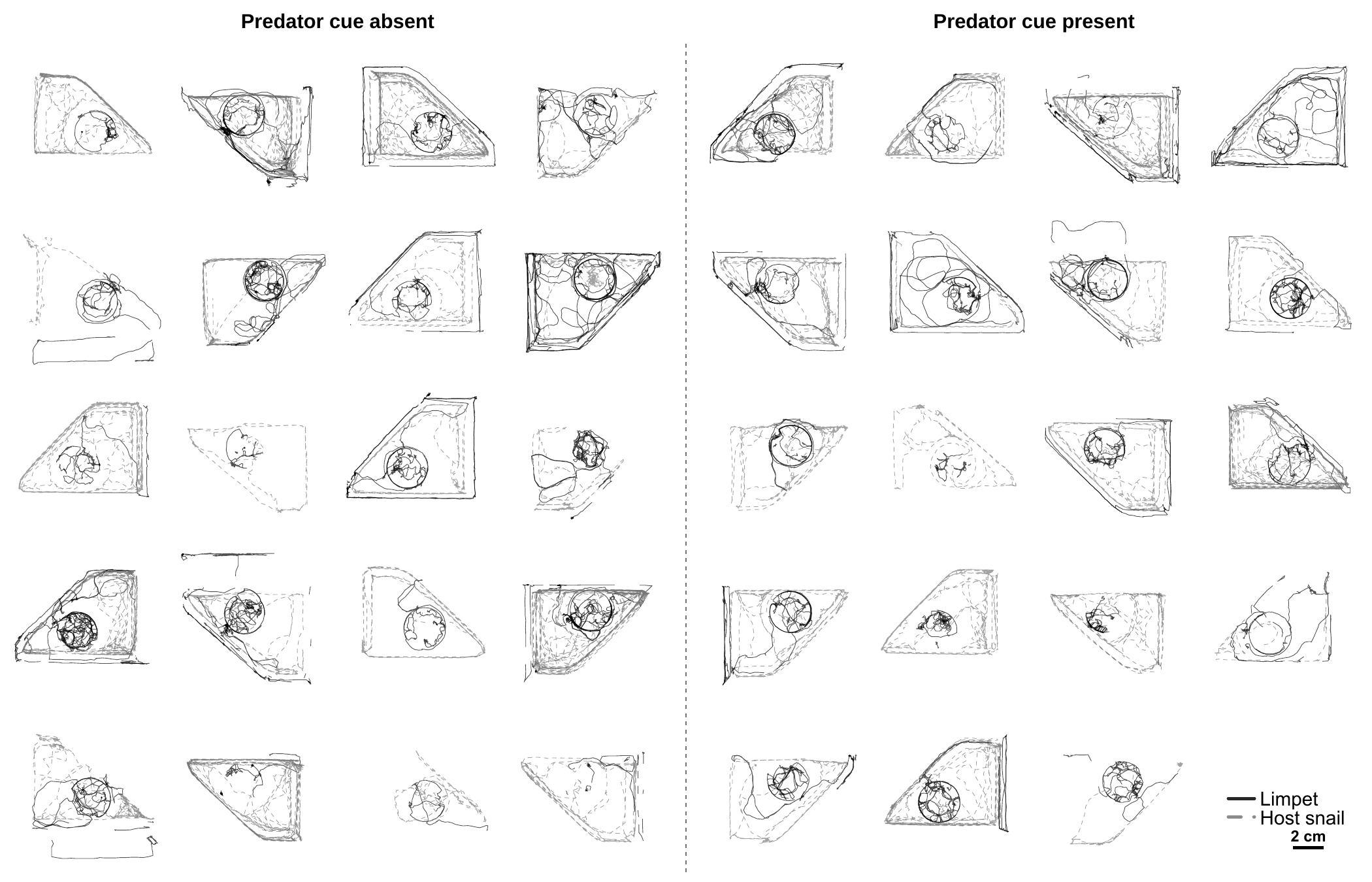}
\caption{All observed host-snail and limpet trajectories by chamber. Black solid lines denote focal limpets and grey dashed lines denote host snails; chambers are grouped by predator-cue treatment. Trajectories are plotted from the processed DeepLabCut-derived coordinates used for the locomotion and paired-trajectory analyses. Lines are broken across missing frames or implausible jumps. This descriptive overview, which was not used as an additional inferential test, places the distance and movement summaries in the context of the underlying tracking data.}\label{fig:all-trajectories}
\end{figure}
\end{landscape}
\clearpage
\begin{figure}[htbp]
\centering
\includegraphics[width=\textwidth,height=0.82\textheight,keepaspectratio]{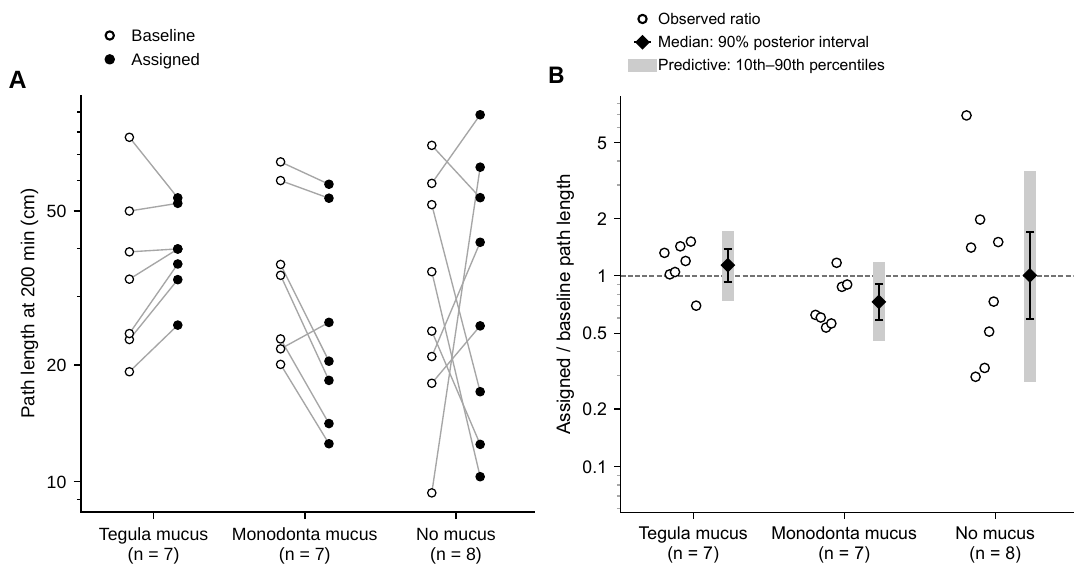}
\caption{\textbf{Mucus-exposure assay at a common observed time point.} (A) Observed cumulative path length at 200 min for 22 limpets with both a no-mucus baseline and an assigned run. Open circles show baseline runs and filled circles show assigned runs; grey lines connect runs from the same individual. Within each treatment, baseline and assigned observations occupy fixed horizontal positions without individual jitter. (B) Observed path-length ratios $R_i$ (assigned run divided by baseline; open circles). Black diamonds show estimates of the population median in each treatment, $\exp(\mu_g)$, summarised by its posterior median. Thin black error bars show central 90\% posterior intervals for that parameter; wide light-grey bars show 10th--90th percentile posterior predictive ranges for $R$ in a new individual. Seven individuals were assigned to \textit{Tegula} mucus, seven to \textit{Monodonta} mucus, and eight to no mucus. Both panels use logarithmic vertical axes.}
\end{figure}

\begin{figure}[htbp]
\centering
\includegraphics[width=\textwidth,height=0.82\textheight,keepaspectratio]{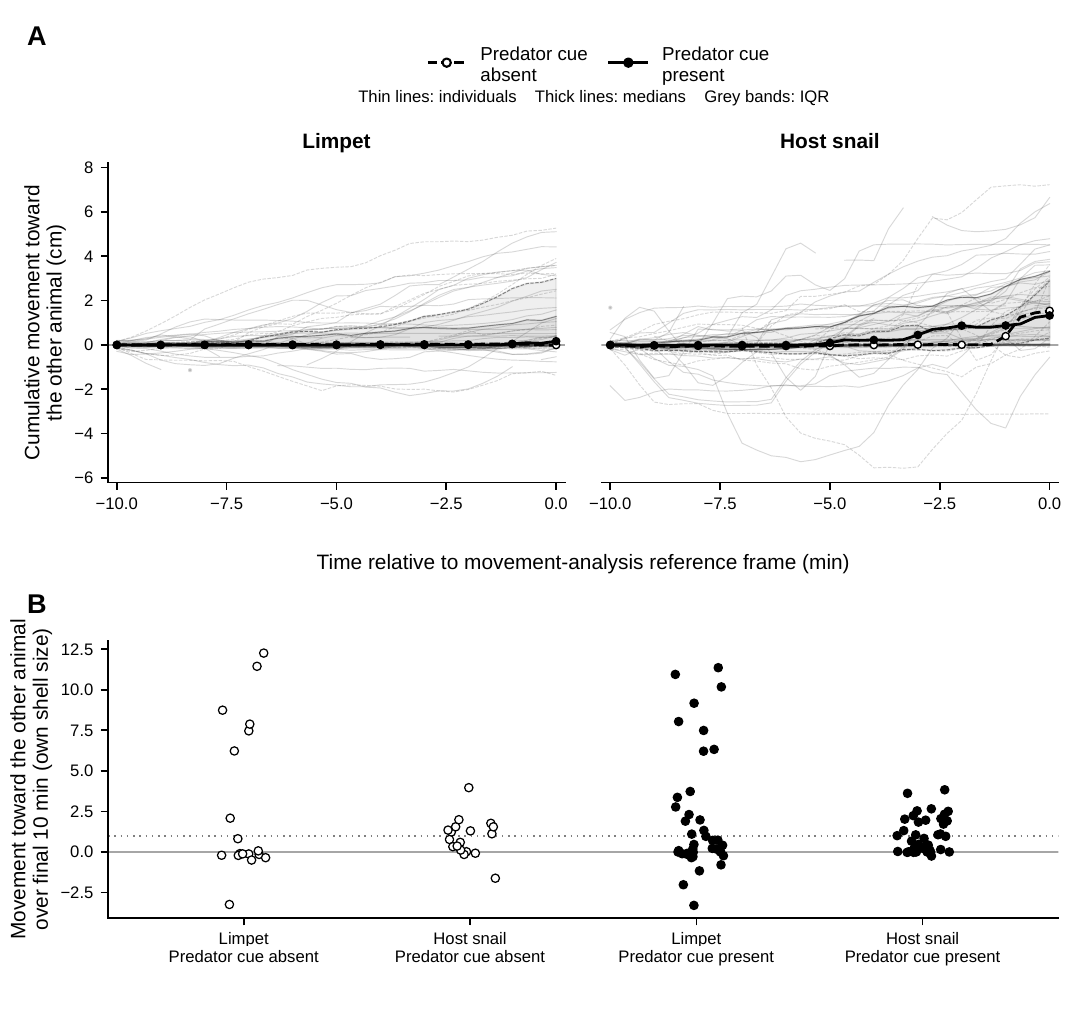}
\caption{\textbf{Movement toward the other animal over a nominal ten-minute window among eventual riders.} Open circles and dashed lines denote predator cue absent; filled circles and solid lines denote predator cue present. (A) Cumulative signed movement toward the other animal in centimetres. Thin transparent lines show individual trajectories, thick black lines and symbols show treatment medians, and grey ribbons show the 25th--75th percentile ranges. Curves are broken across missing paired frames rather than interpolated. Time 0 is the movement-analysis reference frame defined in Section~\ref{sec:paired-trajectories}. Segments ending from $-10$ to 0 min are included, so a segment ending at $-10$ min can begin 20 s earlier. Positive values indicate cumulative projected displacement toward the other animal. (B) Own-size-normalised movement toward the other animal over the same ten-minute window; each point represents one individual component, with horizontal jitter for visibility. Limpet movement is scaled by limpet shell length and host-snail movement by host shell width. The dotted horizontal line indicates one limpet shell length or one host shell width, as appropriate. Both panels use the same 58 eventual riders (18 predator cue absent and 40 predator cue present). Five-minute results in the text provide a shorter-window check.}
\end{figure}

\begin{figure}[htbp]
\centering
\includegraphics[width=\textwidth,height=0.80\textheight,keepaspectratio]{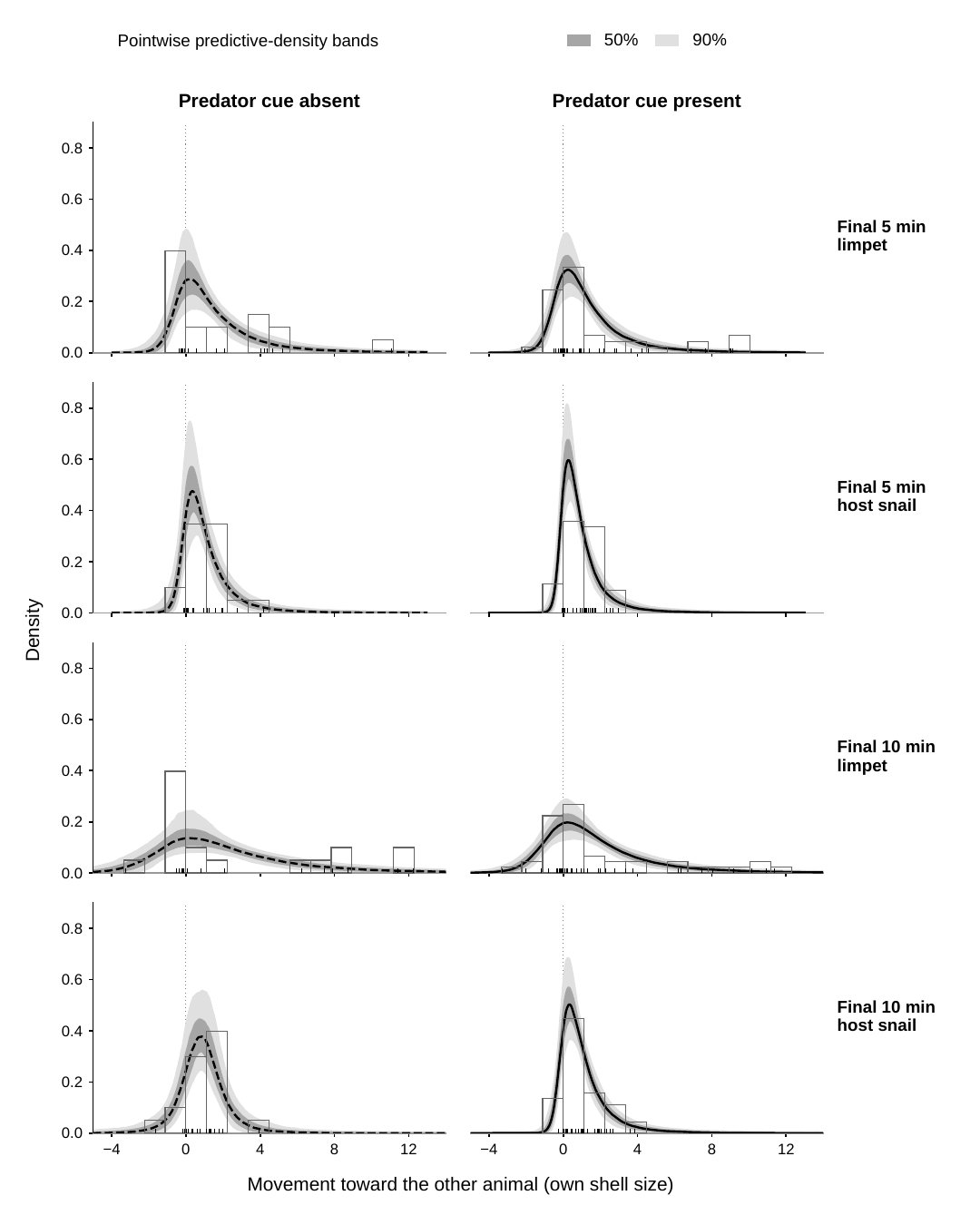}
\caption{\textbf{Posterior predictive checks for skew-$t$ models of movement toward the other animal.} Rows show five-minute and ten-minute windows ending at the movement-analysis reference frame for limpet and host-snail components; columns show predator cue treatments. Outline histograms and rugs show observed own-size-normalised movement values (18 predator-cue-absent and 40 predator-cue-present riders in each window). Black dashed lines (predator cue absent) and solid lines (predator cue present) show posterior median predictive densities. Dark and light grey ribbons show pointwise 50\% and 90\% posterior intervals for the predictive densities, respectively. All eight panels use the same vertical density scale and horizontal movement scale. The horizontal variable is signed projected displacement toward the other animal, scaled by that animal's own shell size, not host--limpet distance.}
\end{figure}

\clearpage

\section{Supplementary movie legends}

\begin{itemize}
\setlength\itemsep{0.4em}
\item \textbf{Supplementary Movie S1.} \path{MovieS1_all_chamber_trajectories_CURRENT.gif}. Fading-trail animation of all visible host-snail and limpet trajectories by chamber during the overnight riding assay. Limpet trajectories are shown as black solid trails with filled-circle endpoints; host-snail trajectories are shown as grey dashed trails with open-circle endpoints. The predator-cue-labelled upper panels allow comparison of locations where limpet and host trails converge or overlap as the recording progresses. The lower panels show host--limpet separation in centimetres over time, with a moving time marker. Spatial overlap is a descriptive feature of these fading trails; simultaneous contact, riding, trail following, and predator-cue-treatment effects each require separate evidence. The animation provides a qualitative overview of how limpets and host snails move through each chamber over time, complementing the distance-closure and movement-component analyses.
\end{itemize}